\documentclass[useAMS,usenatbib]{mn2e}
\usepackage{graphicx}
\usepackage{natbib}
\usepackage{tablefootnote}
\usepackage{amsmath}
\usepackage{amssymb}
\pdfoutput=1
\title[The Nature of the Stingray Nebula from Radio Observations]{The Nature of the Stingray Nebula from Radio Observations}
\author[Harvey-Smith et al.]{Lisa Harvey-Smith$^{1,4,5}$\thanks{Email: lisa.harvey-smith@csiro.au}, 
Jennifer A Hardwick$^{1}$, 
Orsola De Marco$^{2}$,\newauthor
Mudumba Parthasarathy$^{3}$, 
Ioannis Gonidakis$^{1}$, 
Shaila Akhter$^{4,1}$, \newauthor
Maria Cunningham$^{4}$
 and James A Green$^{1}$\\
$^{1}${CSIRO Astronomy and Space Science, Australia Telescope National Facility, PO Box 76, Epping 1710, Australia}\\
$^{2}${Department of Physics, Macquarie University, Sydney NSW 2109, Australia}\\
$^{3}${Indian Institute of Astrophysics,  Bangalore 560034, India}\\
$^{4}${School of Physics, University of New South Wales, NSW 2052, Australia}\\
$^{5}${Western Sydney University, Locked Bag 1797, Penrith, NSW 2751, Australia}}
\begin{document}

\date{}

\pagerange{\pageref{firstpage}--\pageref{lastpage}} \pubyear{2017}

\maketitle

\label{firstpage}

\begin{abstract}

We have analysed the full suite of Australia Telescope Compact Array data for the Stingray planetary nebula. Data were taken in the 4- to 23-GHz range of radio frequencies between 1991 and 2016. The radio flux density of the nebula generally declined during that period, but between 2013 and 2016 it shows signs of halting that decline. We produced the first spatially resolved radio images of the Stingray nebula from data taken in 2005. A ring structure, which appears to be associated with the ring seen in HST images, was visible. In addition, we found a narrow extension to the radio emission towards the eastern and western edges of the nebula. We derived the emission measure of the nebula - this decreased between 1992 and 2011, suggesting that the nebula is undergoing recombination. The radio spectral index is broadly consistent with a free-free emission mechanism, however a single data point hints that a steeper spectral index has possibly emerged since 2013, which could indicate the presence of synchrotron emission. If a non-thermal component component has emerged, such as one associated with a region that is launching a jet or outflow, we predict that it would intensify in the years to come.
\end{abstract}

\begin{keywords}
stars : AGB and post-AGB -- planetary nebulae : general -- radio continuum - individual : SAO 244567 (Hen 3-1357) -- stars : evolution
\end{keywords}

\section{Introduction}
The transition between the asymptotic giant branch (AGB) and the white dwarf phases of a star's life encompasses some of the most complex physics in stellar evolution. Stars undergoing such a transition are difficult to model in one dimension, because they are hydrodynamic systems undergoing pulsations and significant mass loss.
From observations we know that the mass-loss rate of stars on the upper part of the AGB is stupendous \citep[e.g. up to 10$^{-3}$M$_{\odot}$~year$^{-1}$;][]{1999A&A...351..559V} and that the transition between the AGB and post-AGB phases is rapid. 


To add complexity to this picture, approximately 20 per cent of all post-AGB stars also undergo a last helium shell flash, also known as a ``born-again" event \citep{Schoenberner1979,Iben1983} when the post-AGB star suffers a last helium shell flash, expands and temporarily returns to the AGB.
Examples of born-again events are Abell~30 and Abell~78, Sakurai's Object and V605~Aql \citep{Jacoby1981,Duerbeck1997,Clayton2006}. Although the behaviour of born-again events seems to be generally well understood, one or two cases suggest that things are not as simple as they seem. For example, the detection of copious amounts of neon and a carbon to oxygen ratio C/O$<$1 in the ejecta of Abell~30 and Abell~58 \citep{Wesson2003,Wesson2008} is at odds with the theory \citep{Schoenberner1983}. The object discussed in this paper, with its complex behaviour, presents even more cause for doubt.


The subject of this paper is SAO 244567 (CPD -59 69 26 = Hen 3-1357 = IRAS 17119-5926), a post-AGB star that has evolved into the young planetary nebula (PN) over the past three decades \citep{Parthasarathy1993,Bobrowsky1998}. \citet{Parthasarathy1995} classified SAO 244567 as a hot, post-AGB B-type supergiant on the basis of a 1971 spectrum. By 1990$-$1992 the strong [OIII] nebular emission lines \citep{Parthasarathy1993,Parthasarathy1995} indicated that the stellar temperature had increased. Recently \citet{Reindl2014} analysed a 1988 spectrum deriving an effective temperature of 38~kK, confirming that the temperature was indeed increasing rapidly between the early 1970s and late 1980s. Furthermore, \citet{Reindl2017} showed that while in 2002 the star had heated further to 60kK, by 2015 it had actually cooled down to 50kK. They therefore deduced that the post-AGB star had been through a born-again event. That being said, the temperature and luminosity evolution of the star are at odds with any known models of the born again event \citep{Reindl2017}.

 The morphologies of collimated PNe and pre-PNe that result from AGB mass loss cannot be explained (yet) with single-star theory \citep{Soker2006,Nordhaus2007,GarciaSegura2014}, whilst they may be explained as a result of binary interactions \citep[see][for a review]{DeMarco2009b}. The morphology of the Stingray nebula could therefore betray a binary origin. If this were the case, however, one might suspect that the binary interaction may also have interfered with the evolution of the star in some complex way, explaining at least part of its behaviour. If a binary interaction caused the mass loss that dictated the departure from the AGB approximately 1000 years ago, it would be difficult to explain how the same binary may be causing the complex behaviour we have been witnessing in the last 40 years.


Radio-frequency observations can complement knowledge obtained from optical images and spectra. Free-free emission mirrors the ionised gas and is a sensitive meter of the nebular response to the changes in stellar temperature. Non-thermal emission may instead show that magnetic processes are taking place. Free-free radio emission from the Stingray nebula was detected in 1991 by \citet{Parthasarathy1993}, who used the Australia Telescope Compact Array (ATCA) to study the nebula at 6 and 3~cm. 
\citet{2008MNRAS.386.1404U}  studied the radio morphology of the Stingray nebula with the ATCA in March 2000 and August 2002 at 1.5 and 21~cm. The free-free emission from the planetary nebula was dimming, at odds with the expected behaviour of a young planetary nebula which is being increasingly ionised by its heating central star. 

In this paper, we supplement the information gathered from published radio studies of the Stingray nebula with new ATCA data from 2005, 2013 and 2016 at a range of cm-wavelengths to further constrain the nebular behaviour. In Section~\ref{sec:observations} we present the observations taken with the ATCA and discuss our data processing techniques. Section~\ref{sec:results} presents radio continuum images of the Stingray nebula, radio flux densities, emission measure estimates and radio spectral energy distribution and spectral index analysis. We also provide upper limits for non-detections of H$\alpha$, water maser and ammonia spectral lines. In Section~\ref{sec:discussion} we compare radio and optical images and discuss how the radio observations of the Stingray nebula add to our picture of this interesting stage of stellar evolution.

\section{Observations}
\label{sec:observations}

We observed the Stingray nebula around SAO 244567 (RA 17:16:21.071, Dec $-$59:29:23.64, J2000) on 8th August 2013, 24th$-$26th December 2013 and 13th$-$14th March 2016 using the ATCA. We also used archival ATCA data of this object from March 19th$-$22nd 2000, 24th August 2002, 22nd$-$24th June 2005. The August 2013 observations were set up using the calibration source PKS 1929-293 to set the array delays and antenna gains. The flux calibrator PKS 1934-638 was used to set the flux scale and calibrate polarisation leakage terms. PKS 1718-649 was observed every 13 minutes as a phase calibrator.  The December 2013 observations used PKS 1934-638 as a flux and polarisation leakage calibrator and PKS 1657-56 as a phase calibrator. In 2016 we used PKS 1934-638 to set the flux scale and polarisation leakage and PKS 1718-649 to calibrate the phases. 

Table~\ref{table1} provides details of all the observations including the observing date, central frequency, observing bandwidth, number of spectral channels, integration time, array configuration, maximum baseline (as a proxy of angular resolution) and project code. The data from 1991 were published by \citet{Parthasarathy1993}, 2000 and 2002 data were published by \citet{2008MNRAS.386.1404U} and August 2013 data were previously published by \citet{Cerrigone17}. Data from 2005 (project code C1431) were centred at 18.769 GHz (250 kHz resolution);  21.38478 GHz (1 MHz resolution) and 23.694 GHz (60 kHz resolution), which are the rest frequencies of H$_{70}\alpha$, H$_{67}\alpha$, and NH$_3$($J=1-1$), respectively and are previously unpublished. Data from August \& December 2013 and March 2016 were obtained by the authors in Director's Discretionary Time. 

\begin{table*}
\centering
\begin{minipage}{160mm}
\caption[]{Detailed description of the radio observations of the Stingray nebula taken with the Australia Telescope Compact Array that are used in this work. Here we quote the date of observation, observing frequency in megahertz, bandwidth in megahertz and number of spectral channels, total on-source integration time in minutes, array configuration, maximum antenna baseline in metres and ATCA project code. Observations marked P93 were previously published by \citet{Parthasarathy1993}, those marked U08 were published by \citet{2008MNRAS.386.1404U} and those marked C17 were previously published by \citet{Cerrigone17}}.
\label{table1}
\begin{tabular}{@{}ccccccccc@{}}
\hline
Date  & Central Frequency & BW(MHz) & Integration  & Array & Maximum &  Project  & Notes \\
(MMYYYY) &   (MHz) & /$\#$chan & time (mins) & Configuration\footnote{https://www.narrabri.atnf.csiro.au/operations/array\_configurations/configurations.html} & Baseline (m) & Code & \\
\hline
031991 & 4790 & 128/32 & 16 & OLD6K & $\sim$6000 &  C086 & P93\\
041991 & 8640 & 128/32 & 8 & 6A & 5939 &  C086 & P93 \\
\hline
032000 & 4800 & 128/32  & 138 & 6D & 5878 & C858 & U08 \\
032000 & 8640 & 128/32 & 138 & 6D & 5878 & C858 & U08 \\
\hline
082002 & 1384  & 128/32 & 172  & 6C & 6000 & C858 & U08\\
082002 & 1472  &128/32 & 172 & 6C & 6000 & C858 & U08\\
082002 & 2368  &128/32 & 172 & 6C & 6000 & C858 & U08\\
082002 & 4800  &128/32 & 172 & 6C & 6000 & C858 & U08\\
082002 & 8640 &128/32 & 172 & 6C & 6000 & C858 & U08\\
082002 & 16832 &128/32 &  310 & 6C & 6000 &C858 & U08\\
082002 & 18752 &128/32 & 310 & 6C & 6000 & C858 & U08\\
\hline
062005 & 23694  & 16/256 & 435 & 6B & 5969 & C1431 & NH$_{3}$ \\
062005 & 18769  & 32/128 & 481 & 6B & 5969 & C1431 & H$_{70}\alpha$ \\
062005 & 21384  & 64/64 & 917 & 6B & 5969 & C1431 & H$_{67}\alpha$  \\
\hline
082013 & 17000 & 2048/2048 &225 &  750D & 4469 &  CX271& C17 \\
082013 & 19000 & 2048/2048 & 225 &  750D & 4469 &  CX271& C17 \\
\hline
122013 & 2100 & 2048/2048 & 37  & 750B  & 4500 & CX274  & Continuum \\
122013 & 6000 & 2048/2048 & 9  & 750B  & 4500 & CX274  & Continuum \\
122013 & 9500 & 2048/2048 & 9& 750B  & 4500 & CX274  & Continuum \\
122013 & 17000 & 2048/2048 &  17  & 750B  & 4500 & CX274  & Continuum \\
122013 &  22700 & 2048/2048 & 17  & 750B  & 4500 & CX274  & Continuum \\
122013 & 22348 & 288/9216 & 17  & 750B  & 4500 & CX274  & H$_{2}$O\\
122013 & 23244 & 288/9216 & 17  & 750B  & 4500 & CX274  & H$_{2}$O\\
\hline
032016 & 17000 & 2048/2048 & 267& 6B & 5969  & CX271 & Continuum \\
032016 & 19000 & 2048/2048 & 267& 6B & 5969  & CX271 & Continuum \\
\hline
\end{tabular}
\end{minipage}
\end{table*}

We also obtained archival data from observations of the Stingray nebula from the Australia Telescope Online Archive\footnote{http://atoa.atnf.csiro.au} and re-analysed these data. In this paper we present the combined data to study the radio morphology and evolution of this extremely young PN.

\subsection{Data Processing}

Radio data from the ATCA were processed using the {\sc miriad} software package \citep{1995ASPC...77..433S}. Data were loaded using {\sc atlod} and channels known to be corrupted in the correlator (so-called `birdies') were flagged. For the higher frequency bands (above 9~GHz) a correction for atmospheric opacity was also carried out in {\sc atlod}. Radio-frequency interference was flagged using the {\sc pgflag} package using the `SumThreshold' algorithm \citep{2010MNRAS.405..155O}. For each observation we formed a total intensity image using {\sc invert}. Each image was modelled using {\sc clean} and self-calibrated before the final model was convolved with a restoring beam to produce a final total intensity image. To determine the peak flux density {\sc uvfit} was used to fit the visibilities at zero baseline to a 2-dimensional Gaussian. This was performed on all epochs for comparison of flux density.

We re-processed the 18.752, 21.384 and 23.694 GHz data from 2005 and the 22.238 GHz and 23.244 GHz data from December 2013 for consistency in the continuum data reduction and also to search for spectral line emission, which had not previously been published. Raw visibility data were loaded into {\sc miriad}, this time omitting any automated RFI flagging. The flux scale was determined from short observations of PKS 1934-638.  We determined antenna gains, delay terms and passband responses from our phase calibrator sources, (PKS 1718-649 in 2005 and PKS 1253-055 in 2013), which were observed at regular intervals (every $\sim$15mins). These were applied to the Stingray nebula data. Next, we fitted a 3rd order polynomial to the continuum emission and subtracted this from each \textit{u,v}  data set.  The \textit{u,v} data were inspected before and after continuum subtraction, to ensure that the quality of the subtraction was acceptable. The \textit{u,v} data were then Fourier transformed to form image cubes. Total integrated intensity (moment 0) images as well as time- and baseline-averaged spectra, were made. 

\section{Results}
\label{sec:results}

In this section we present images, flux density measurements and spectral indices derived from radio interferometric observations of the Stingray nebula with the ATCA.

\subsection{First resolved radio continuum images of the Stingray nebula} 

\begin{figure*}
\begin{center}
\includegraphics[width=6.4cm, height=8.7cm, angle=270, trim=0cm 1cm 0cm 2cm, clip=true]{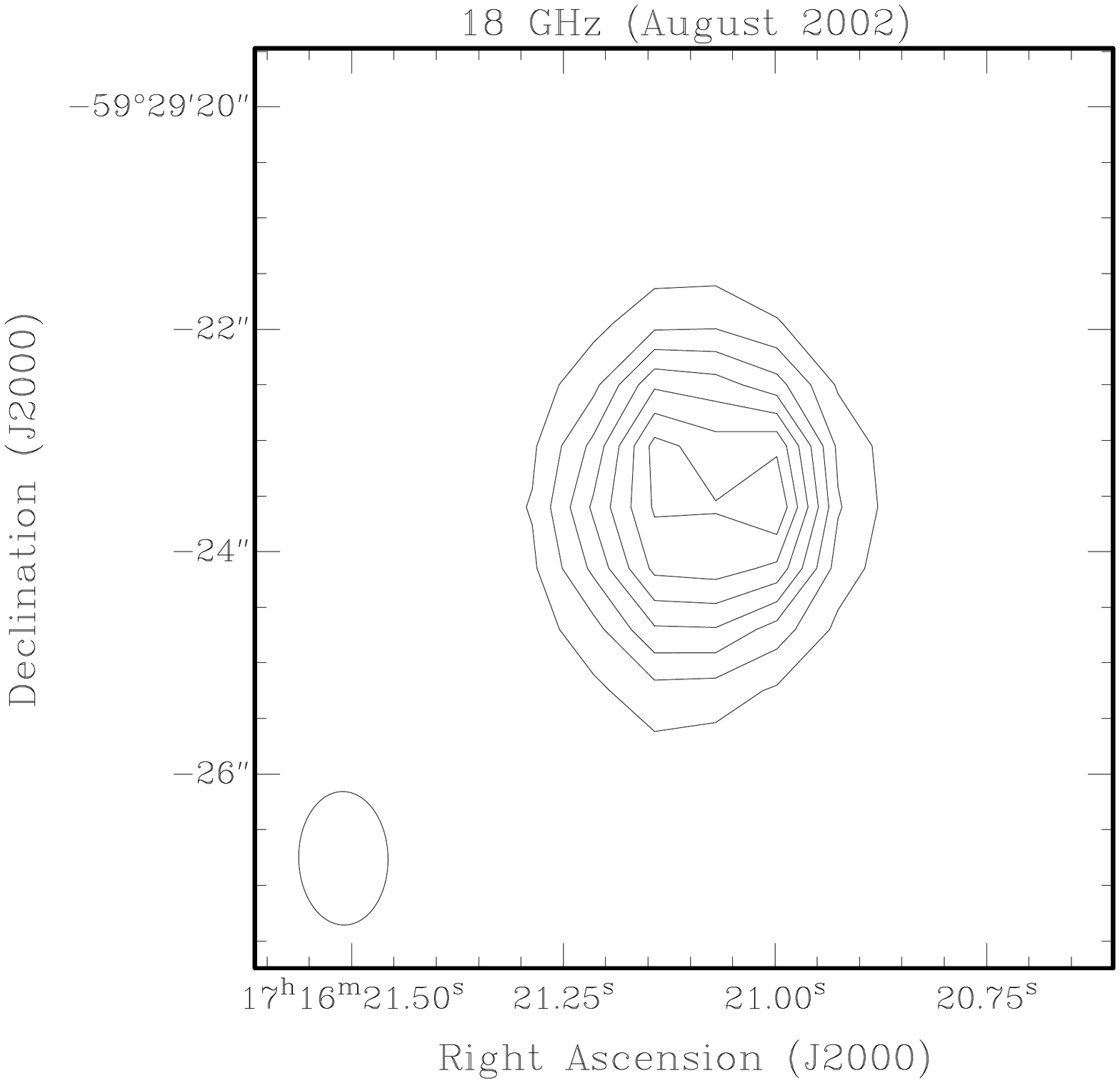}
\includegraphics[width=6.4cm, height=8.7cm, angle=270, trim=1.1cm 1cm 0cm 2cm, clip=true]{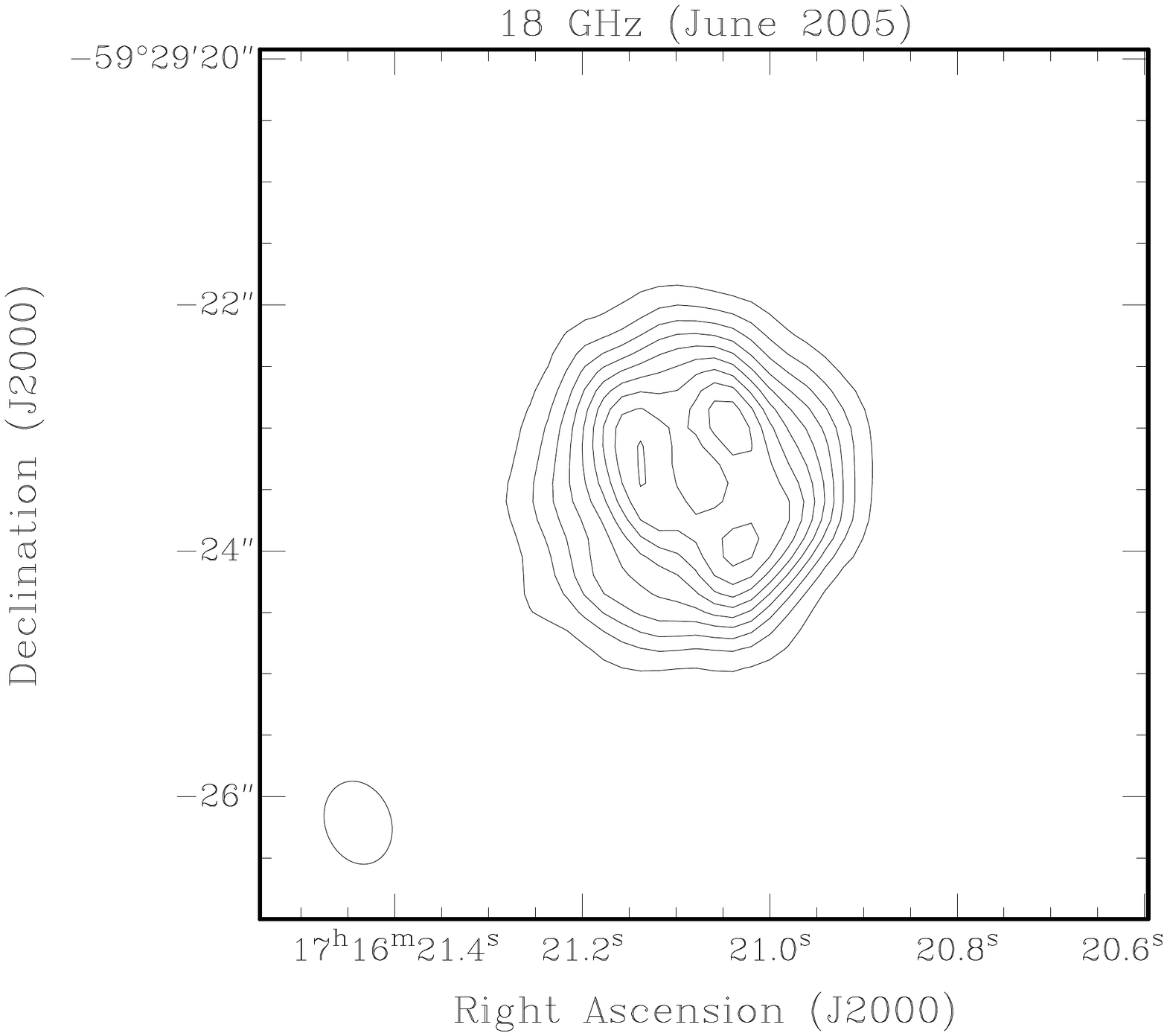}
\includegraphics[width=6.4cm, height=8.7cm, angle=270, trim=1.4cm 1.7cm 0cm 2cm, clip=true]{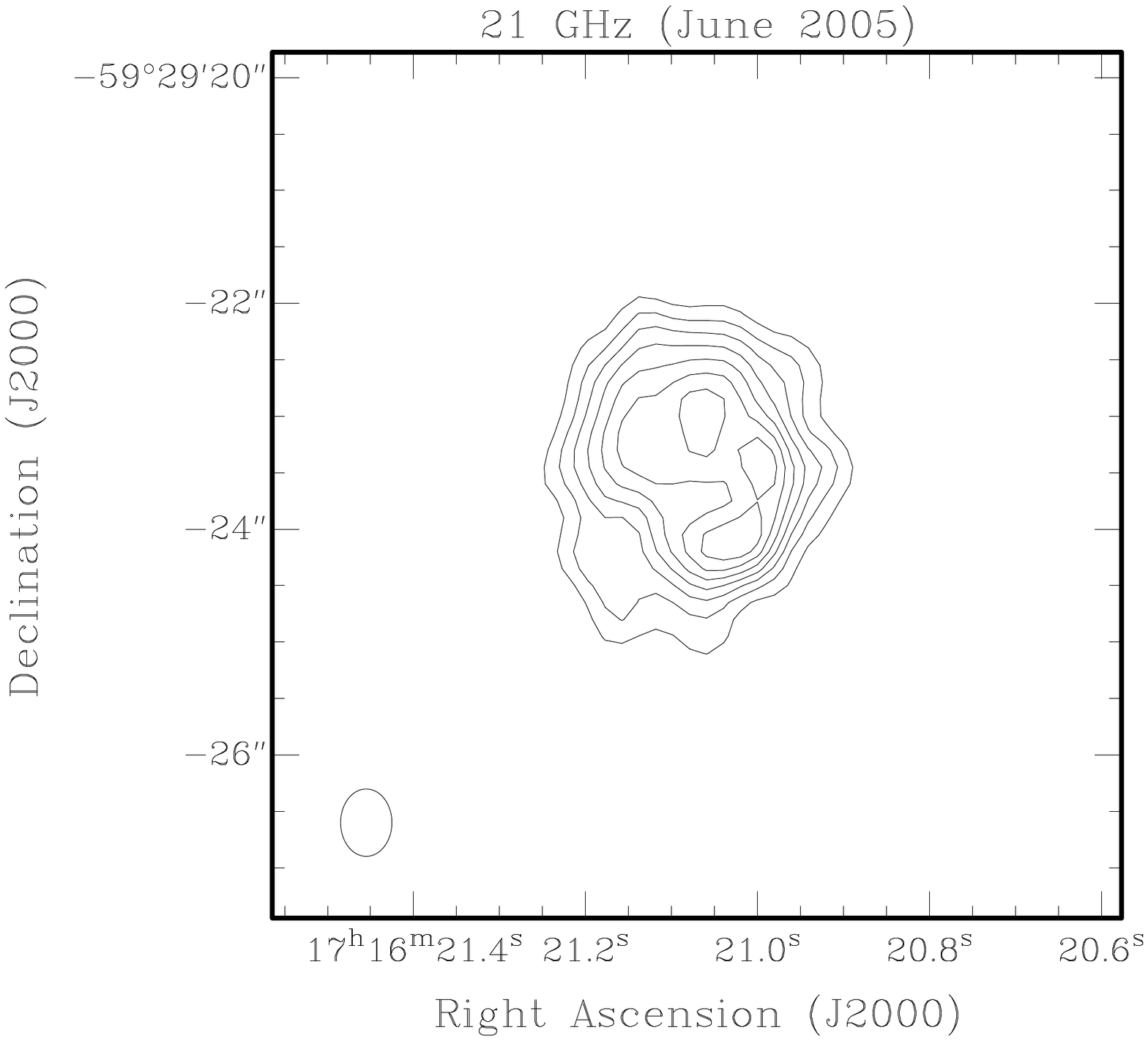}
\includegraphics[width=6.4cm, height=8.7cm, angle=270, trim=0.5cm 1.5cm 0cm 2cm, clip=true]{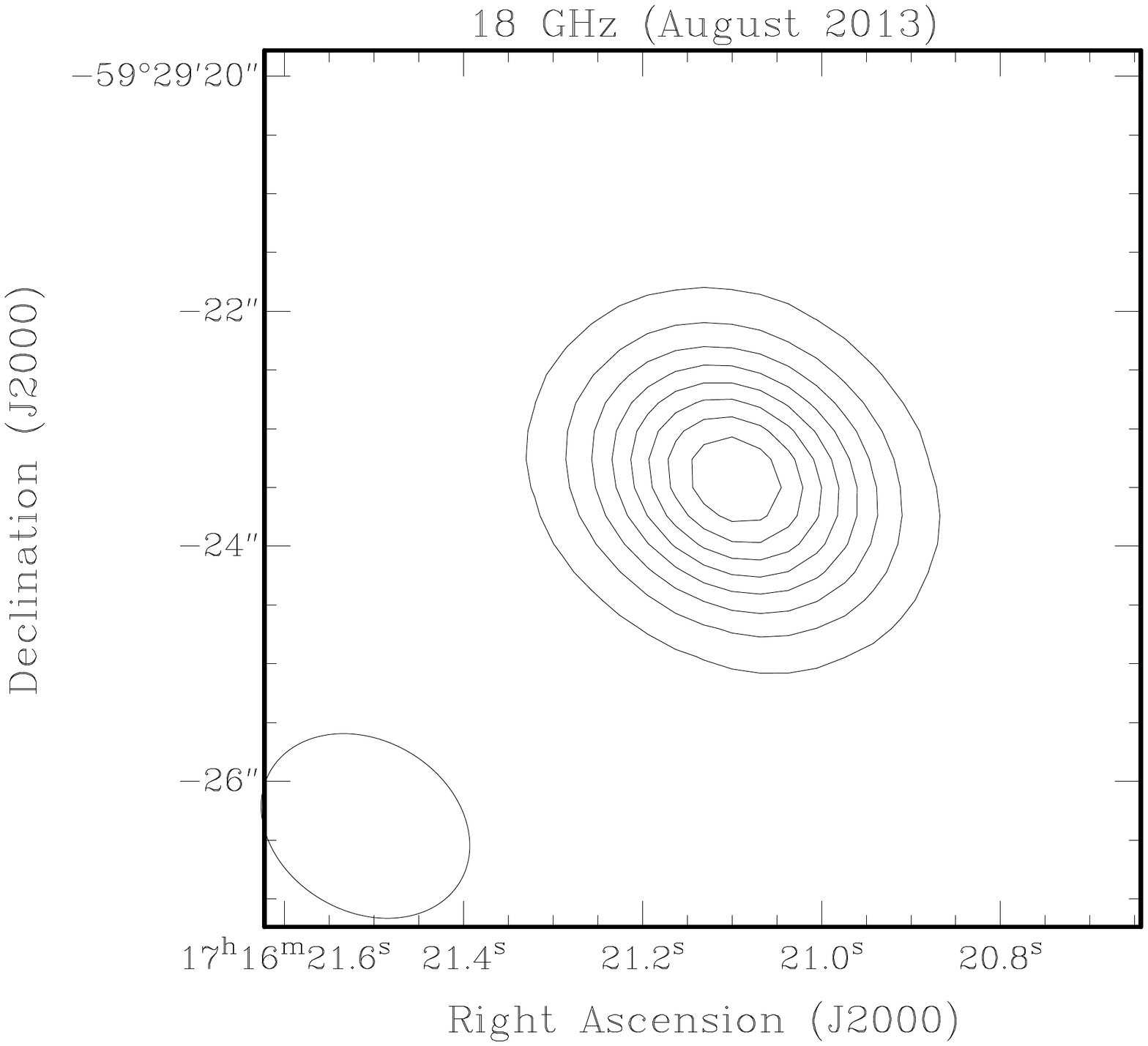}
\includegraphics[width=6.4cm, height=8.7cm, angle=270, trim=0.7cm 1cm 0cm 2cm, clip=true]{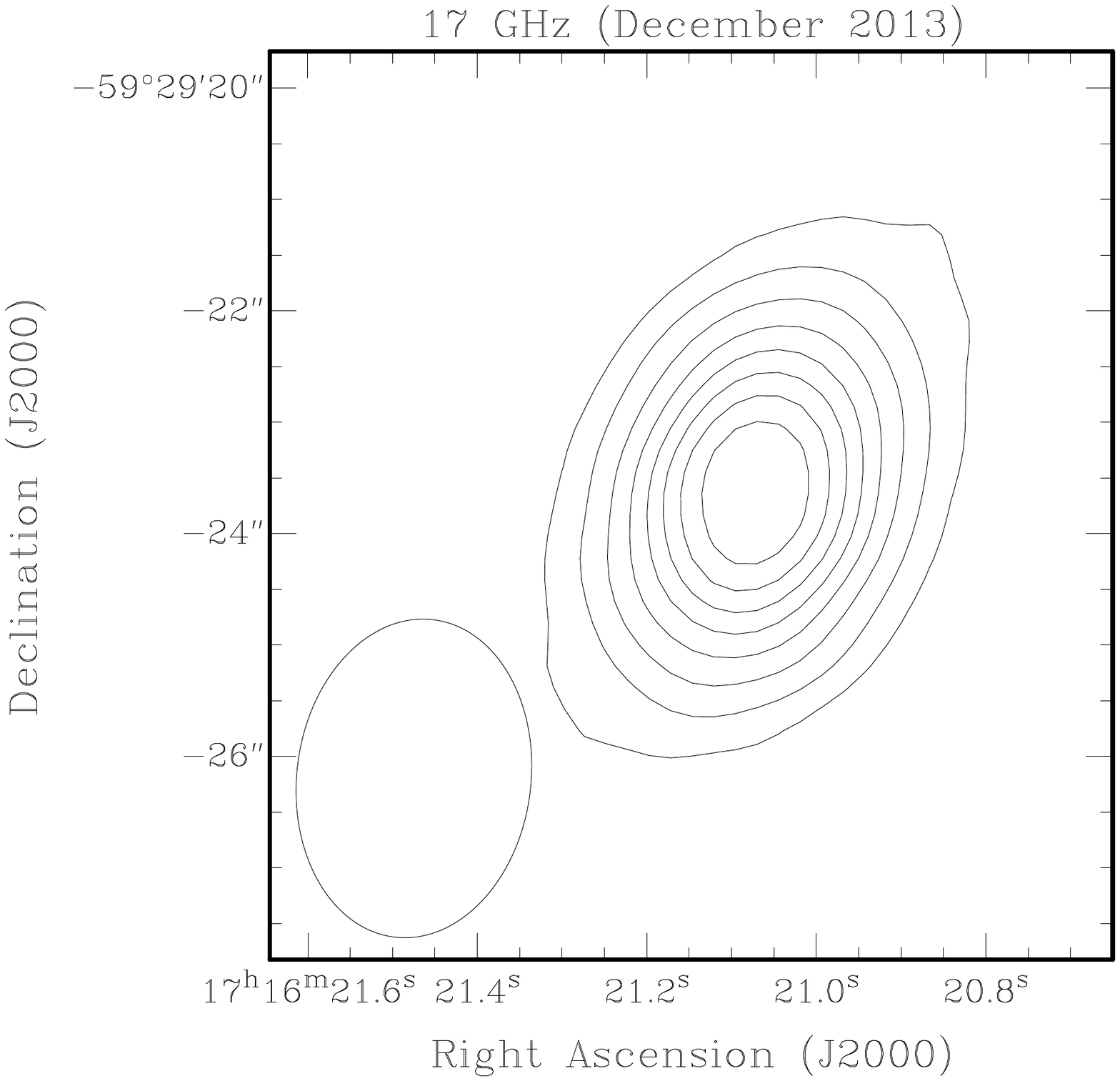}
\includegraphics[width=6.4cm, height=8.7cm, angle=270, trim=0cm 1cm 0cm 2cm, clip=true]{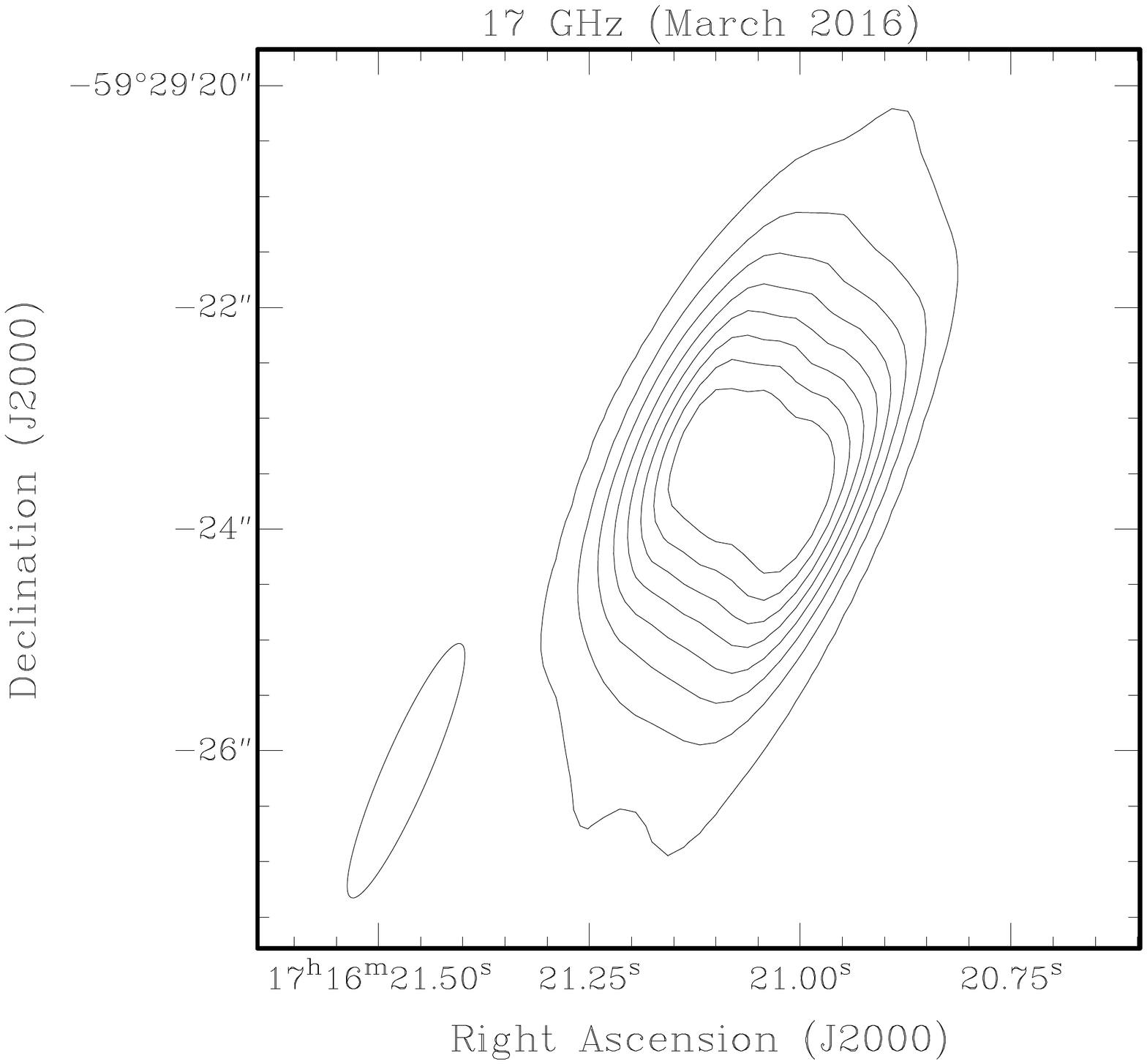}
\caption{Total intensity contour maps of the Stingray nebula observed with the ATCA (top, L-R) at 18.7~GHz in August 2002 and at 18.7~GHz in June 2005 (middle, L-R) at 21.4~GHz in June 2005 and at 18.0~GHz in August 2013 and (bottom, L-R) at 17.0~GHz in December 2013 and 17.0~GHz in March 2016. Contours are from 10\% of the peak to 90\% of the peak value, in increments of 10\%. Ellipses at the bottom-right indicate the size and shape of the synthesised beam of the interferometer. During this time the star is understood to have been undergoing a phase of rapid cooling and expansion, although the nebula itself shows only weak sign of expansion between 2002 and 2016.}
\label{fig1}
\end{center}
\end{figure*}

Figure~\ref{fig1} shows contour maps of the radio emission detected using the ATCA at approximately 17$-$21 GHz, integrated across the full range of observed frequencies in each frequency band. Most of the images are partially (or completely) unresolved, but the 2005 epoch was taken at a sufficiently high angular resolution and long integration time to produce a spatially-resolved image. This shows for the first time the radio morphology of the Stingray nebula. The 2013 and 2016 images also appear slightly resolved but the insufficient angular resolution and \textit{u,v}-coverage of these observations prevents us from making detailed images for those epochs.

Our resolved images of the Stingray nebula confirm the presence of an inner ring of radio emission close to the post-AGB progenitor and an extended near-spherical nebula of radio emission. In the 18-GHz image the ring nebula appears somewhat symmetrical, however at 21-GHz we note evidence for a possible gap in the eastern side of the inner ring and the emergence of two `ears', i.e. a slight extension of the radio emission at either side of the nebula at an angle approximately 45$^{\circ}$ to the image axes. These extensions were not visible in the 18-GHz data, which may be due to the fact that the angular resolution is slightly lower. We discuss the morphology of the planetary nebula and compare this to optical images from the HST in Section~\ref{sec:discussion}.


\subsection{Radio flux density monitoring of the Stingray Nebula}

In order to determine how the emission profile (and the spectral energy distribution) of the nebula has changed over time, we calculated the radio flux density of the Stingray nebula at each observing epoch and gathered all the available radio flux density information from the literature. Table~\ref{table2} shows the results including the date, central observing frequency in MHz and the single-dish equivalent flux density in millijansky calculated by extrapolating the measured radio flux density vs. baseline length back to a zero-baseline position using the {\sc miriad} task {\sc uvfit}.

\begin{table}
\centering
\begin{minipage}{90mm}
\caption[]{Single-dish equivalent flux densities of the Stingray nebula calculated from data gathered with the Australia Telescope Compact Array between 1991 and 2016. Observations marked P93 were previously published by \citet{Parthasarathy1993} and those marked U08 were published by \citet{2008MNRAS.386.1404U}. The observations from August 2013 were previously published by \citet{Cerrigone17} but here we quote the results of our own analysis, for consistency of method.} The remainder of the radio observations are new to this work.
\label{table2}
\begin{tabular}{@{}ccccc@{}}
\hline
Date  & Observing &  Single-dish equivalent &  Notes \\
(MMYYYY) & frequency (MHz) & flux density (mJy)&  \\
\hline
031991 & 4790 & 63.3$\pm$1.8 & P93 \\
041991 & 8640 & 51$\pm$12 & P93 \\
032000 & 4800 & 57.6$\pm$1.7 & U08 \\
032000 & 8640 & 52.0$\pm$1.6   & U08 \\
082002 & 1384 & 36.6$\pm$1.1   &U08 \\
082002 & 2368 & 46.9$\pm$1.8   & U08\\
082002 & 4800 & 48.8$\pm$1.5   & U08 \\
082002 & 8640 & 46.6$\pm$1.4   & U08 \\
082002 & 16832 & 43.8$\pm$2.0 & U08 \\
082002 & 18752 & 42.8$\pm$2.0   & U08 \\
012005 & 18769 & 36.0$\pm$3.6 & -\\
012005 & 21384 & 40.2$\pm$4.0 & -  \\
082013 & 17000 & 26.2$\pm$0.8  & - \\
082013 & 19000 & 26.3$\pm$0.9 & -\\
122013 & 2100 & 25.4$\pm$0.8 & - \\
122013 & 6000 & 30.0$\pm$0.9 & -\\
122013 & 9500 & 29.2$\pm$0.9  & -\\
122013 & 17000 & 23.4$\pm$0.7   & -\\
122013 & 22700 & 20.8$\pm$0.9  & -\\
032016 & 17000 & 25.7$\pm$0.8   & -\\
032016 & 19000 & 25.5$\pm$0.8   & -\\
\hline
\end{tabular}
\end{minipage}
\end{table}

The total uncertainty in the flux density measurements, $\sigma$, comprises two components; the uncertainty in the fit to the \textit{u,v} data, $\sigma_{fit}$, and the uncertainty in the absolute flux scale $\sigma_{cal}$, transferred from the flux calibrator source PKS 1934-638 is given by Equation~\ref{eq1}:

\begin{equation}
\sigma^2=\sigma_{fit}^2+S \sigma_{cal}^2,
\label{eq1}
\end{equation}

\noindent where $S$ is the measured flux density of a source. By comparing two simultaneous independent measurements of the phase calibrator source PKS 1929-293 in the adjacent 17000- and 19000-MHz bands, we estimated the absolute error in the flux scale determination ($\sigma_{cal}$) to be 3\%. For epochs 2005 and earlier (prior to the installation of the Compact Array Broadband Backend) we assume the uncertainty in the absolute flux scale to be 10\% (Jamie Stevens, private communication, 2016). $\sigma_{fit}$ was determined by the {\sc miriad} task {\sc uvfit}.

\begin{figure}
\begin{center}
\includegraphics[width=9cm, trim=0.5cm 7cm 0cm 7cm, clip=true]{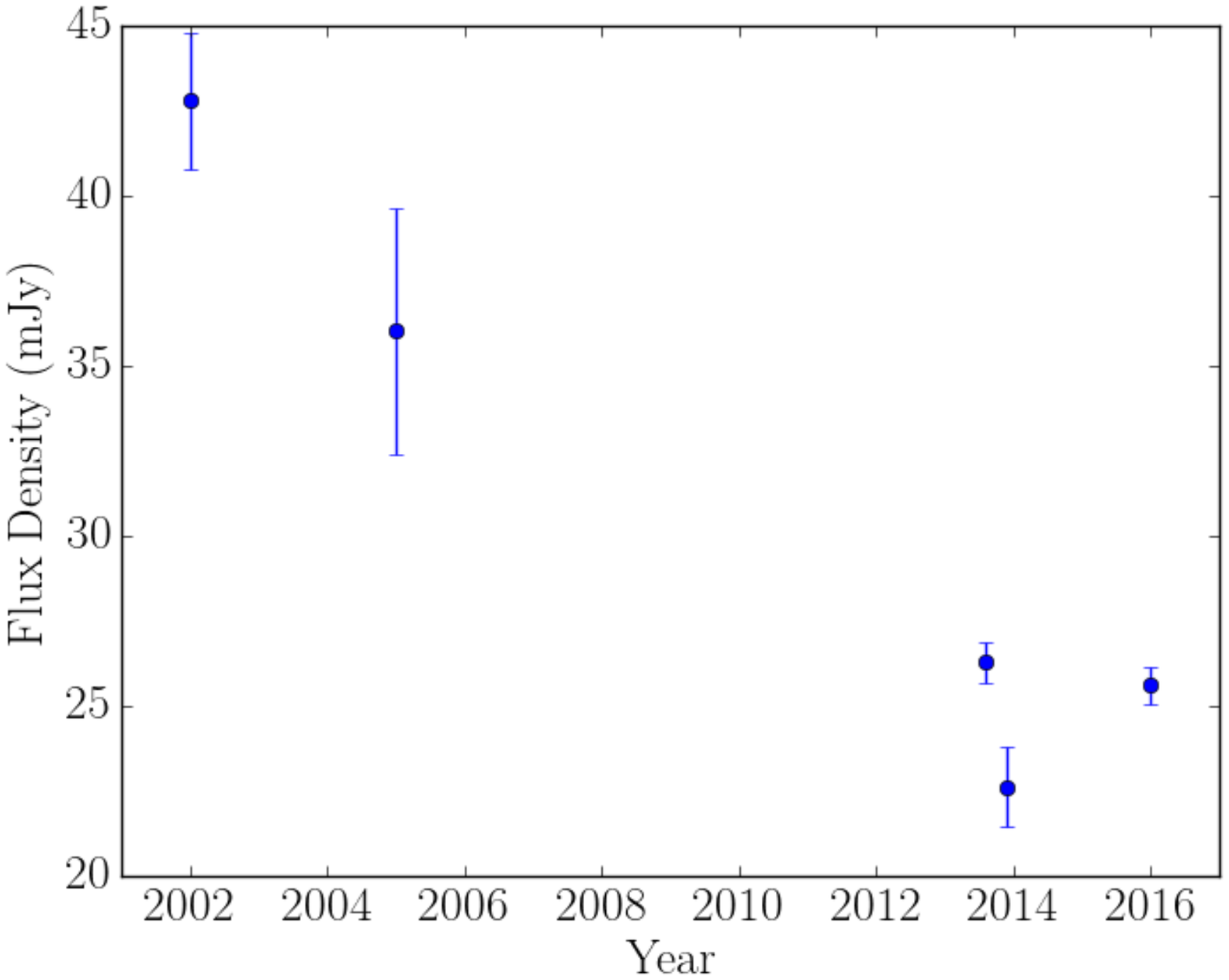}
\caption{The 18.7 GHz flux density of the Stingray nebula measured between 2002 and 2016. Flux density values are either measured directly at 18752 (August 2002) and 18769 (June 2005) or interpolated from measurements in the 17000 and 19000 MHz bands (August 2013 \& March 2016) and the 17000 and 22700 MHz bands (December 2013). The radio flux density has declined steadily over the past 14 years, as the nebula has expanded and recombined and the final data point indicates a possible flattening or increase in the radio emission between 2014 and 2016.}
\label{fig2}
\end{center}
\end{figure}

Figure~\ref{fig2} shows the change in the 18.7~GHz continuum flux density of the Stingray nebula between 2002 and 2016. The continuum flux density decreased steadily between 2002 and 2014, increasing slightly by 2016. 

\subsection{The emission measure of the Stingray Nebula}
We can also use the variation in radio flux density to estimate the ionisation status of the Stingray nebula. According to \cite{1999acfp.book.....L} the emission measure in an optically thin nebula is given by Equation~\ref{eq2}:

\begin{equation}
EM \approx \frac{S_{\nu} {T_e}^{1/2}}{2 \theta^2},
\label{eq2}
\end{equation}

\noindent where EM is the emission measure in pc cm$^{-6}$, S$_{\nu}$ is the radio flux density in jansky at frequency $\nu$ in hertz, T$_e$ is the electron temperature in kelvin and $\theta$ is the apparent angular extent of the source in degrees.

Using Equation~\ref{eq2}, an estimate of $T_e$=11,500~K for both 1992 and 2011 from \citep{2013AstL...39..201A}, estimates of $\theta$=1.5 and 1.7 arc seconds for the angular size of the nebula in 1992 and 2011 respectively based on measurements from \citep{Parthasarathy1993} and this work, and radio flux density measurements at 8640~MHz from those years (this paper, extrapolated between measurements assuming a linear time-variation in radio flux density) we determine the emission measure, EM, to be 1.5$\times$10$^7$pc~cm$^{-6}$ in 1992 and 6.7$\times$10$^6$pc~cm$^{-6}$ in 2011. The emission measure of the planetary nebula has reduced very rapidly. This reduction in emission measure is due to the expansion of the nebula and recombination as the central star has cooled and its ionising radiation flux has decreased.

\subsection{Spectral Energy Distribution}

\begin{figure} 
\begin{center}
\includegraphics[width=9cm]{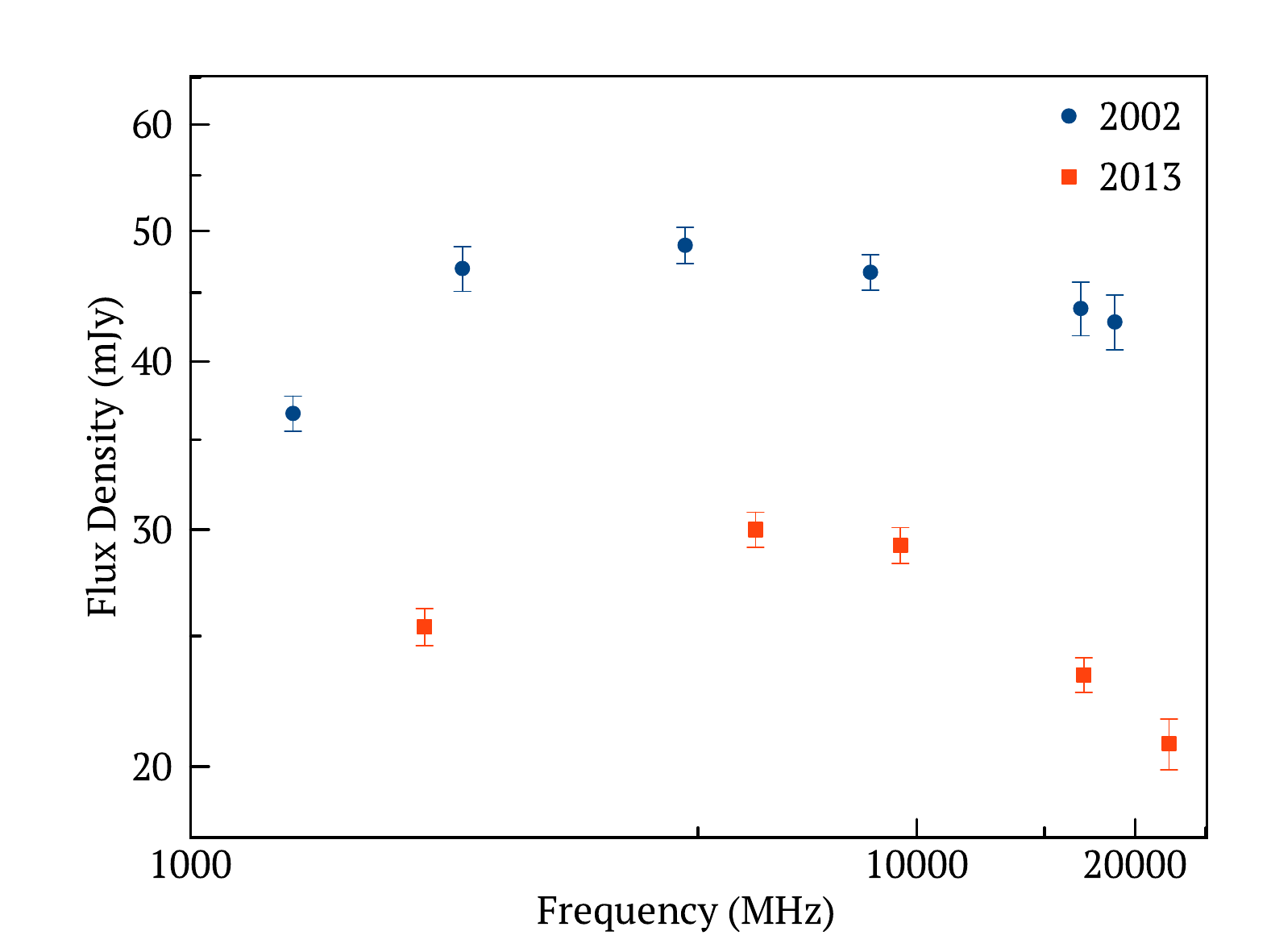}
\caption{Spectral energy distribution of the Stingray nebula in 2002 and 2013, measured using the ATCA. Axes are plotted on a log-log scale and the spectral steepening is apparent between these epochs.}
\label{fig3}
\end{center}
\end{figure}

The spectral energy distribution (SED) is a plot of energy, brightness or flux density versus frequency. The SED of a planetary nebula enables us to probe the physical conditions and emission processes within the nebula. For two of our observing epochs (August 2002 and December 2013) the wide range of observing frequencies enabled us to study how the SED of this source has varied as the Stingray nebula has evolved. Figure~\ref{fig3} shows the spectral energy distribution in 2002 and 2013. The flux density of the source has diminished quite significantly in the cm-radio band since 2002, probably due to the expansion and/or recombination of the nebula. During the same time, the turnover frequency has increased from 2712 MHz in 2002 to 6693 MHz in 2013. At the same time, the spectral index of the nebula has steepened above this peak frequency.


The slope of the SED at frequencies higher than the turnover frequency gives information on the emission mechanism within the optically thin region of the nebula. The spectral index, $\alpha$, is given by Equation~\ref{eq3}. 

\begin{equation}
\alpha(\nu)=\frac{\partial\log S\nu}{\partial\log\nu}.
\label{eq3}
\end{equation}
 
In planetary nebulae the spectral index is usually $\alpha$=$-$0.1 as this is dominated by thermal radio emission \citep{1991ApJ...378..599A}. We calculated the spectral index above the turnover frequency in 2002 and 2013, finding $\alpha$=$-$0.1$\pm$0.1 in August 2002 and $\alpha$=$-$0.4$\pm$0.1 in December 2013. Interestingly, the 2013 epoch is inconsistent with a purely thermal (free-free) radio emission mechanism, and may suggest the emergence of non-thermal (synchrotron) component in recent years.

\begin{figure}
\begin{center}
\includegraphics[width=9cm, trim=0.5cm 7cm 0cm 7cm, clip=true]{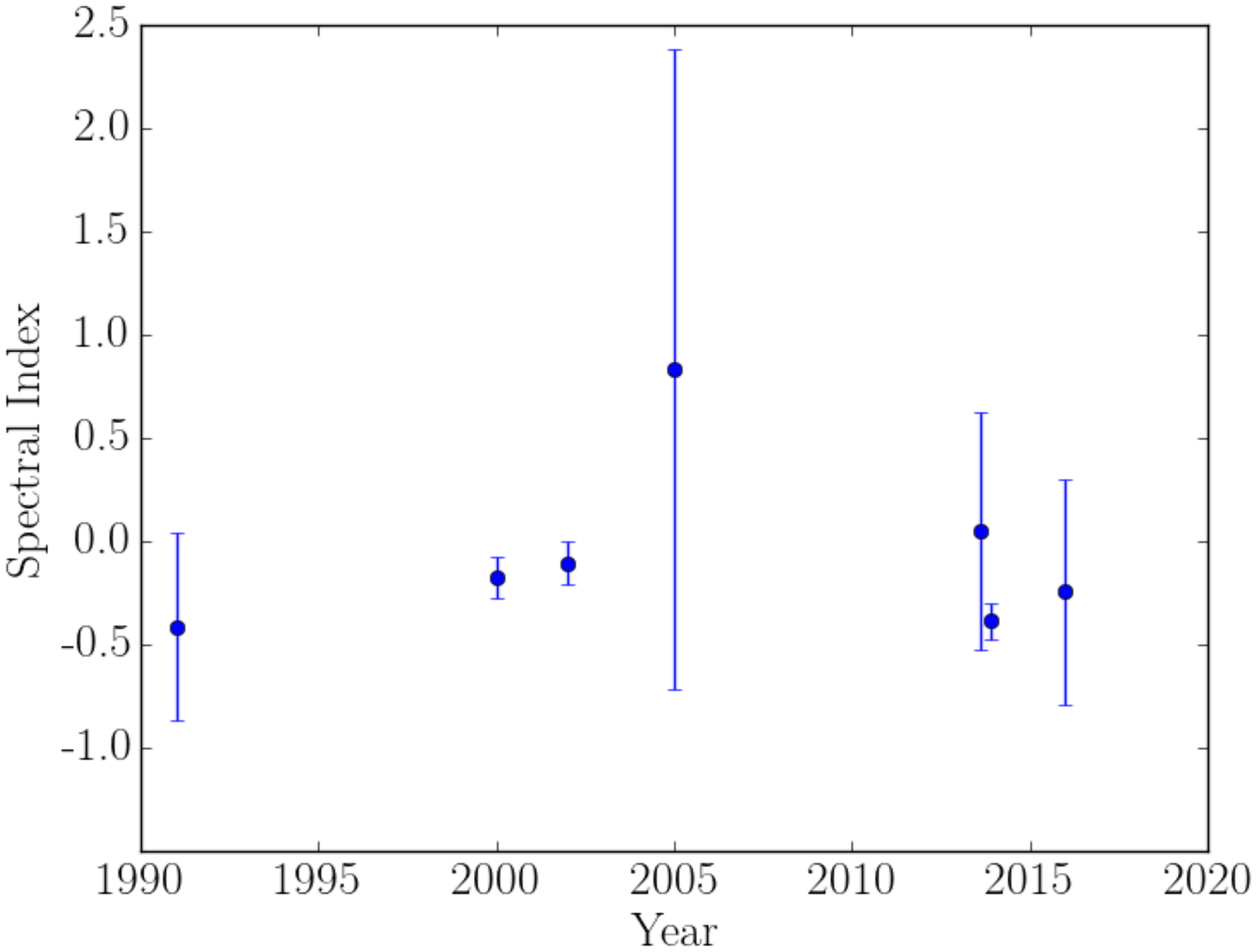}
\caption{The spectral index of the Stingray nebula in the optically-thin part of the spectrum, estimated using ATCA observations listed in Table~\ref{table1}. Data from August 2002 and December 2013 are derived from data that span an octave of frequency bandwidth whereas the other epochs are estimates of the optically-thin spectral index derived from flux density measurements in two independent frequency bands. Results from all epochs are consistent with $\alpha$=$-$0.1, as expected for free-free emission from an ionised planetary nebula apart from the December 2013 data, which indicate a steeper spectral index close to $\alpha$=$-$0.4. This could possibly indicate the emergence of non-thermal (synchrotron) radio emission.}
\label{fig4}
\end{center}
\end{figure}

To bolster our spectral index data set, we also considered less well-sampled SEDs (i.e. those with only two measured frequencies). In Figure~\ref{fig4} we present all our spectral index data for the Stingray nebula, taken in 1991, 2000, 2003, 2005, 2013 and 2016. Uncertainties were very high in the epochs with only two measured frequency bands, which limits our ability to distinguish the possible emergence of non-thermal emission. In particular, the spectral index in 2005 is highly uncertain because flux density measurements in only three closely spaced frequency bands were available. However, the 2013 data (which are well sampled, with seven independent frequency measurements spread out across more than 21 MHz in total bandwidth) are pointing to the emergence of non-thermal component. Future studies of this nebula should be conducted with at least a similarly broad and well-sampled frequency coverage to ensure adequate sensitivity to changes in the spectral index.

\subsection{Spectral line non-detections}

High spectral-resolution radio observations of the Stingray nebula in 2005 and 2013 sought to detect emission from H$_{67}\alpha$, H$_{70}\alpha$, H$_2$O and NH$_3$, which could reveal information about the ionisation conditions and possibly about outflows (accompanied by H$_2$O maser emission) or dense gas (from NH$_3$) in the region of the nebula.

However, inspection of our calibrated and continuum-subtracted spectral line data revealed no emission above the rms noise level of 0.97, 0.44, 2.2 mJy/beam (18.752, 21.384 and 23.694 GHz in 2005) and 3.7 and 3.2 mJy/beam (22.238 and 23.132 GHz in 2013).

Many post-AGB stars go through a mass-loss phase which is associated with water maser emission that are generated by collisional pumping of a water-molecule-rich environment surrounding the star. The region was previously searched for evidence of water masers by \cite{2009A&A...505..217S} using the Parkes radio telescope, however no emission was detected above the rms noise level of 170 mJy. The non-detection of water masers from the Stingray nebula using 2013 spectral line data suggests that the central star is not currently undergoing significant mass-loss by way of a collimated outflow entrained within a molecular environment. This finding is consistent with the observation that the nebula has been largely cleared of its dense molecular material (no visible NH${_3}$) and ionised by the central star over the past two decades of activity.

\section{Discussion}
\label{sec:discussion}

\begin{figure}
\begin{center}
\includegraphics[width=7.6cm, trim=0cm 6.5cm 0.5cm 6cm, clip=true]{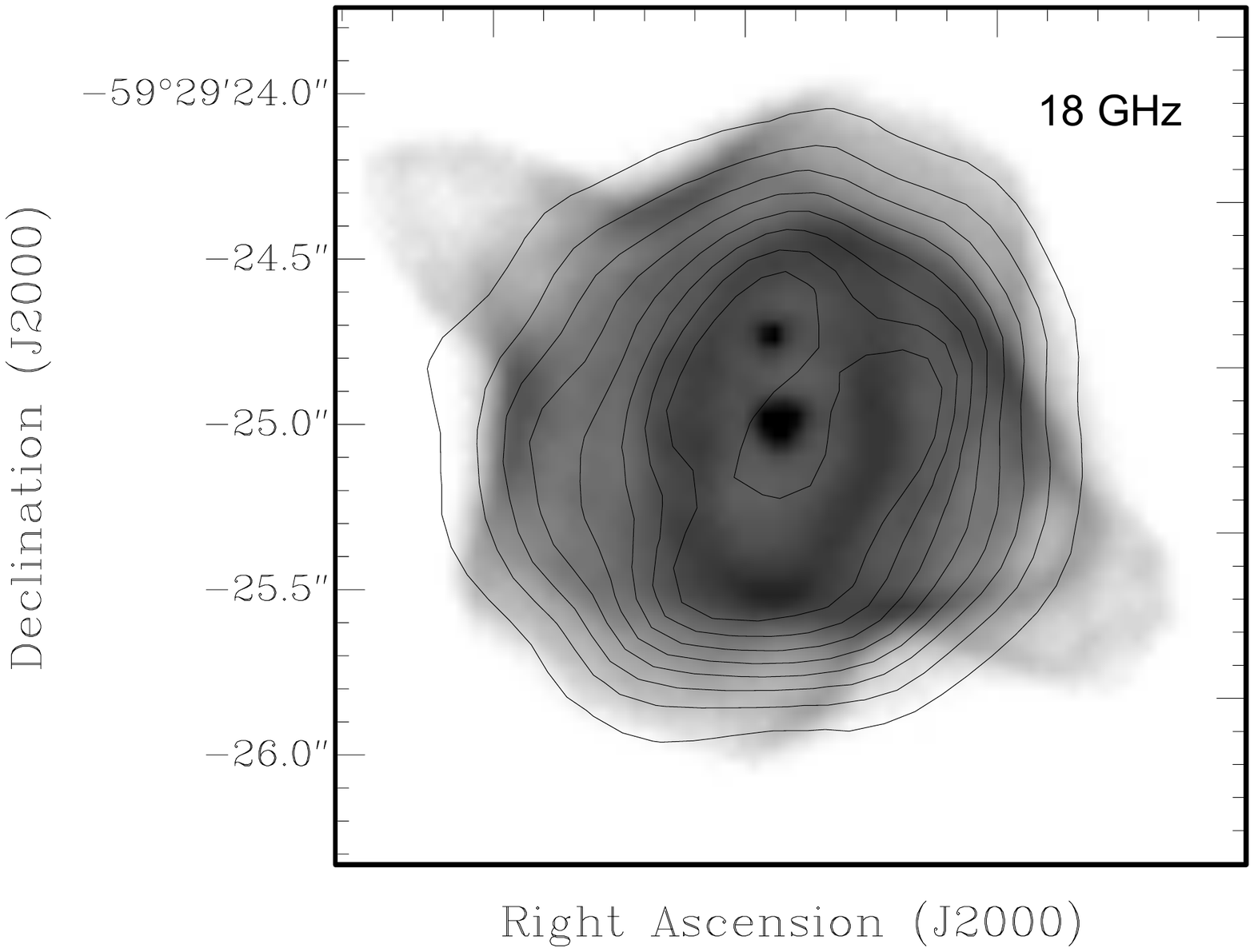}
\includegraphics[width=7.6cm, trim=0cm 6.5cm 0.5cm 6cm, clip=true]{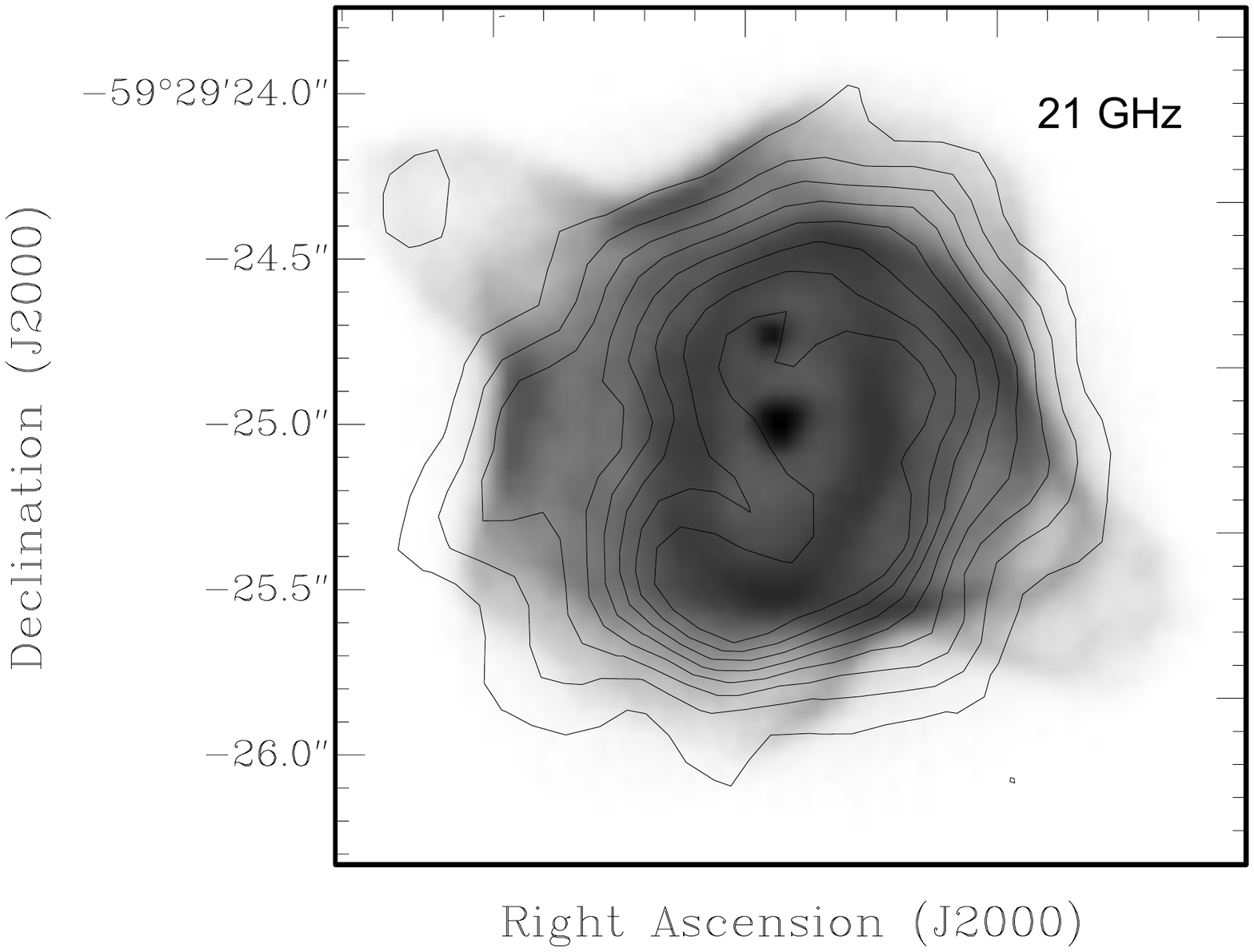}
\includegraphics[width=7.6cm, trim=0.7cm 8.5cm 2.0cm 6cm, clip=true]{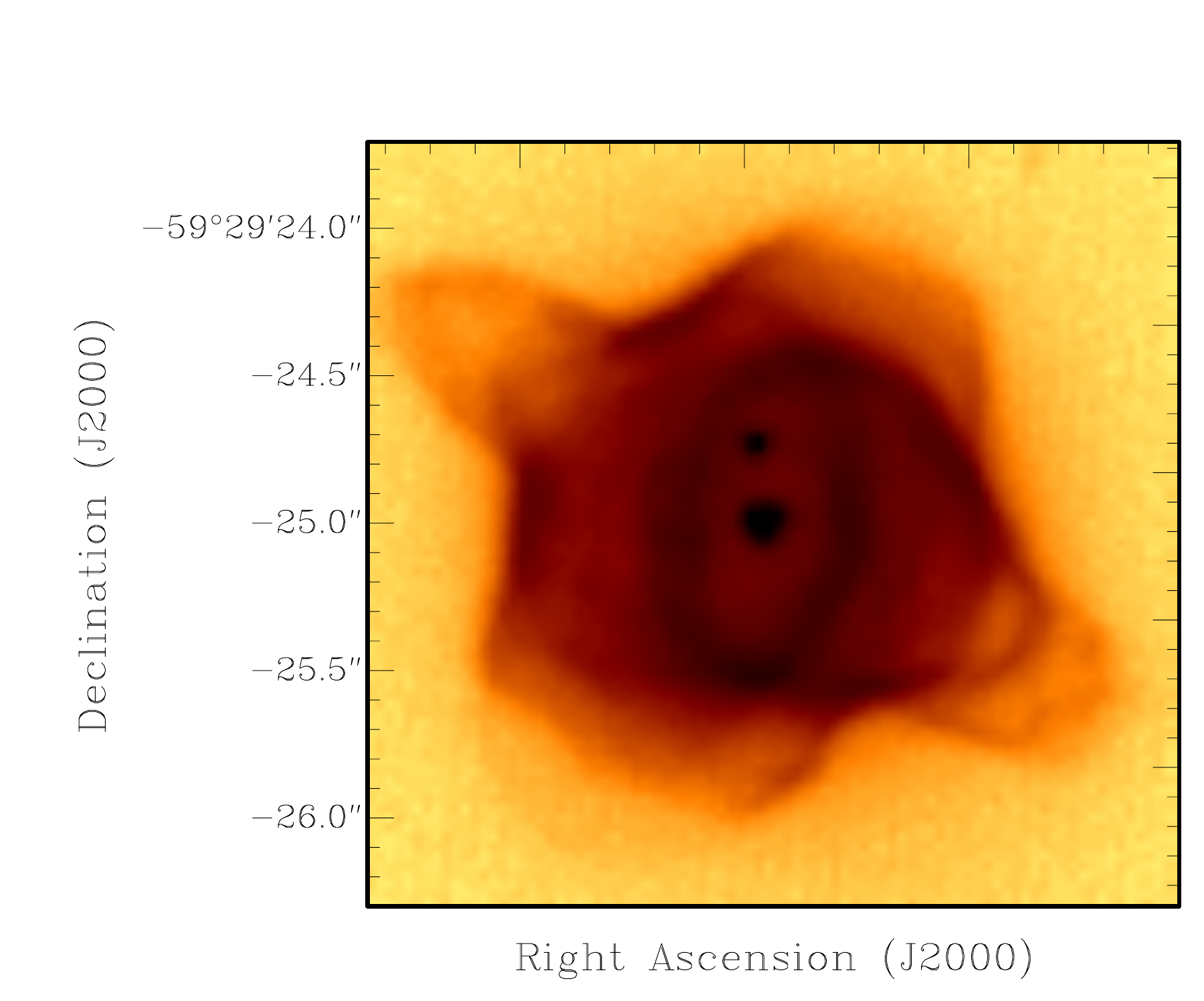}
\caption{Top and middle: Radio images (contours) of the Stingray nebula taken with the ATCA in 2005 at 18~GHz (top) and 21~GHz (middle) overlaid onto optical images from the HST taken in 2001 (greyscale). Note the `ears' visible in the 21~GHz radio contours, which are aligned roughly with the inner collimated outflow marked on Figure~\ref{fig:labels}. Contours are from 10\% of the peak radio flux to 90\% of the peak radio flux, in increments of 10\%. Bottom: False colour image of the Stingray nebula from the HST taken in 2001, contrast chosen to emphasise ring features. The orientation of the images is rotated with respect to those in Figures~\ref{fig1} and \ref{fig:labels}.}
\label{figHST1}
\end{center}
\end{figure}

To compare the radio morphology of the Stingray nebula with the ultra-violet/optical/infra-red morphology, we obtained Space Telescope Imaging Spectrograph images taken in August 2001 from the HST online archives\footnote{https://archive.stsci.edu/hst/search.php}. The HST image is shown in Figure~\ref{figHST1}, overplotted with radio contours from the 2005 ATCA 18~GHz and 21~GHz observations.

\begin{figure}
\begin{center}
\includegraphics[width=6.7cm, angle=270, trim=0cm 2cm 0cm 2cm, clip=true]{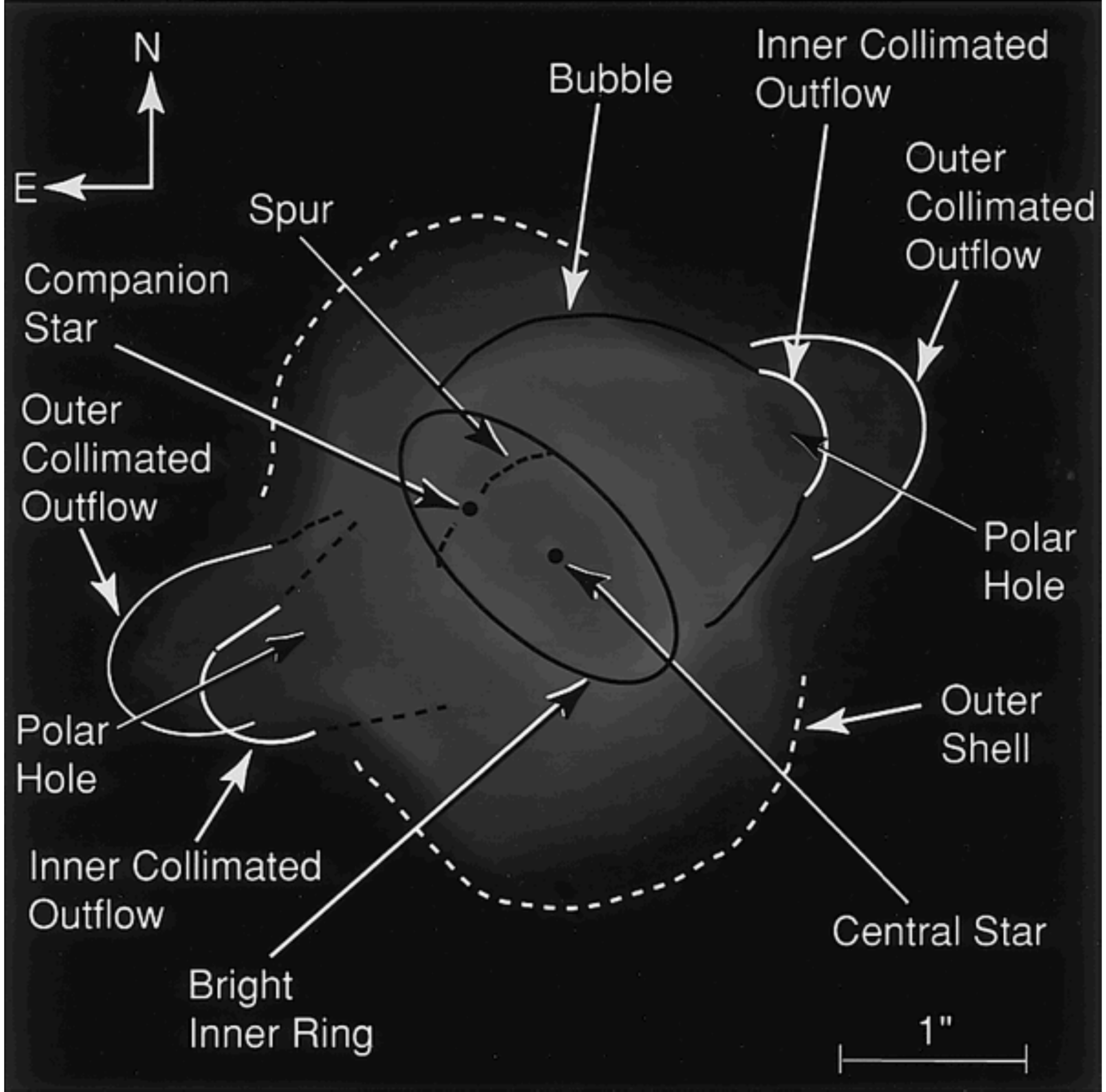}
\caption{Labeled optical image of the Stingray nebula, reproduced with permission from \citet{Bobrowski99}. Note that the sky projection is rotated by 90$^{\circ}$ to the orientation in Figure~\ref{figHST1}.}
\label{fig:labels} 
\end{center}
\end{figure}

A bright ring of emission is visible in both radio images - coinciding with the ring seen in the optical image from the HST. Our observation of a radio ring confirms the conclusions of \citet{2008MNRAS.386.1404U}, who predicted the presence of this feature using modelling analysis of their unresolved 8-GHz ATCA \textit{u,v}-data. The shape of the radio nebula appears slightly different at 18- and 21-GHz. The 18-GHz ATCA contours shown in Figure~\ref{figHST1} show a reasonably regular ring with a small gap from a section of the north-western side of the ring. In our 21-GHz radio image we see the appearance of two `ears' - i.e. bipolar extensions of the nebula, the position angle of which is close to the angle of the `inner collimated outflow' seen in HST images \citep[structures labeled in Figure~\ref{fig:labels}, reproduced with permission from][]{Bobrowski99}. 

The presence of `ears' has been ascribed to the emergence of jets in PNe \citep{Tsebrenko2013}. Combined with the possible emergence of non-thermal radio synchrotron emission, the `ears' may point to a magnetically-driven collimated outflow emerging from the star. Since the radio `ears' are closely aligned with the inner outflow noted by \citet{Bobrowski99}, our understanding of the picture requires further investigation by way of well-sampled radio SEDs taken now and into the future. We plan to continue this monitoring work using the ATCA.

\citet{Cerrigone17} found that variability in a small sample of young PNe was probably linked to a non-thermal component of radio emission. If the Stingray Nebula is indeed growing a non-thermal component then we might perhaps expect the total radio flux density of the nebula to increase in the coming years. Indeed, we have already seen a hint of this behaviour in the recent flattening of the total radio flux density of the Stingray nebula. It will be interesting to continue to watch this phenomenon in the coming years.

We find clear evidence that the nebula is recombining (i.e. the emission measure is declining) and the nebula is expanding, which is consistent with the hypothesis that SAO 244567 has undergone a late thermal pulse and is now on its way back to AGB \citep[i.e. `born again';][]{Reindl2017}.

Recently \citet{Otsuka17} made a detailed analysis of the high-resolution spectrum of the nebula. Based on the chemical composition, the concluded that the initial mass of the central star was 1.0$-$1.5~M$_{\odot}$ None of the present evolutionary models are able to satisfy the observed parameters and rapid evolution. Our radio images weakly indicate that the radio nebula may have expanded between 1991 and 2013, although the observations had different angular resolutions and sensitivities so the picture is not entirely clear. Further high-resolution radio imaging of the nebula may enable us to derive the expansion parallax and hence a distance to the object. 

As we continue to monitor the star and its nebula over the coming years, we will gain a clearer picture of the evolutionary pathway that this star is travelling and thus, learn more about the later stages of stellar evolution from AGB star$\rightarrow$post-AGB star$\rightarrow$planetary nebula and the mechanism of the so-called `born again' phase.


\section{Conclusions} 

We have produced the first resolved radio images of the Stingray nebula, from data taken in 2005. A ring structure is seen, which coincides spatially with the ring seen in HST images. Extensions to the radio emission (`ears'), particularly prominent on the eastern edge of the nebula, are also seen. The emission measure of the nebula decreased between 1992 and 2011, suggesting that the nebula is expanding and/or the plasma is undergoing recombination.

While binary interactions can offer additional ways to explain complex behaviour, in the case of the Stingray nebula it is not obvious how a binary interaction may be used to create scenarios that are in line with the observations. An interaction such as a Roche lobe overflow, possibly followed by the establishment of a common envelope, could lead to the growth of a complex morphology, but it would have likely resulted at the same time in a much hotter central star with a temperature behaviour in line with envelope removal. On the other hand, a common envelope interaction would not naturally explain the cooling observed after the heating.

We have monitored the radio flux density of the Stingray nebula at several radio frequencies between 1991 and 2016, finding a steady decline over that period at 18.7~GHz but noting a slight increase between 2013 and 2016. This increase is consistent with the possible emergence of a non-thermal component of radio emission. The radio spectral index is largely consistent with a free-free emission mechanism, but hints of a change in the radio spectral index to values below $-$0.1 (a value typical for planetary nebula) since 2013 may indicate the intensification of a non-thermal component such as radio synchrotron emission. This result is low in significance, since although the error margin on the data point is small, there is only one such data point on which to base this conclusion. However, if the emergence of non-thermal emission is in line with the emergence of a jet, we predict that future observations should show an intensification of the non-thermal component. Further monitoring observations using the ATCA are therefore planned.

\section*{Acknowledgments}

The Australia Telescope Compact Array is part of the Australia Telescope National Facility which is funded by the Australian Government for operation as a National Facility managed by CSIRO. This paper includes archived data obtained through the Australia Telescope Online Archive (http://atoa.atnf.csiro.au). The authors acknowledge Jasmine Chuawiwat and Michelle Thomes for their assistance with the observations table, Mark Wieringa for his help with {\sc miriad} and Jamie Stevens for his advice on ATCA data reduction. The authors are grateful to the reviewer, Grazia Umana, for her insightful comments that improved our paper.

\bibliographystyle{mn2e}
\bibliography{bibliography} 
\label{lastpage}
\end{document}